\documentclass[11pt]{article} 
\begin{document} 
  \vspace*{1.1cm} 
  \begin{center} 
  {\Large \bf The dipolar zero-modes of Einstein action: \\
  An informal summary with some new issues} 
  \end{center} 

  \begin{center} 
  \vskip 10pt 
  Giovanni Modanese\footnote{e-mail address: 
  giovanni.modanese@unibz.it}
  \vskip 5pt
  {\it California Institute for Physics and Astrophysics \\
  366 Cambridge Ave., Palo Alto, CA 94306}
  \vskip 5pt
  and
  \vskip 5pt
  {\it University of Bolzano -- Industrial Engineering \\
  Via Sernesi 1, 39100 Bolzano, Italy}
  
\vskip 10pt

To appear in the proceedings of 
``Gravitation and Cosmology: from the Hubble Radius to the
Planck Scale. A Symposium in Celebration of the 80th
Birthday of Jean-Pierre Vigier", World Scientific Editor.

\end{center} 

  \baselineskip=.175in 
    
\begin{abstract}
We recall the main features of metric vacuum fluctuations
 which have the global property $\int d^4x \sqrt{g(x)}
 R(x) = 0$, even though $R(x) \neq 0$ locally. 
 We stress that these fluctuations could mediate an
 anomalous coupling between the gravitational field and
 coherent matter. Some
 new issues are discussed: (1) these fluctuations still
 imply that $\langle T_{\mu \nu}(x) \rangle =0$; (2)
 they are not extrema of the action; (3) for finite
 duration, their volume in phase space is not zero;
 (4) vacuum fluctuations of this kind are not allowed in
 QED; (5) their null-action property is a
 nonperturbative feature; (6) any {\it real} pure
 e.m.\ field generates zero-modes of this kind, too,
 up to terms of order $G^2$.

\end{abstract}

 In this note we describe a set of gravitational field 
 configurations, called ``dipolar zero modes", which have 
 not been considered earlier in the literature. They give an 
 exactly null contribution to the pure Einstein action and 
 can thus represent large vacuum fluctuations in the 
 quantized theory of gravity. 
  
 The basic idea behind dipolar fluctuations was discussed 
 for the first time in our earlier work on stability of 
 Euclidean quantum gravity \cite{prd59}; the Lorentzian 
 case was treated in \cite{plb460}. This year we made the 
 first explicit computations and we were able to set some 
 lower bounds on the strength of the fluctuations 
 \cite{npb,prd}. Also we gave for the first time in Ref.s 
 \cite{npb,prd}: (1) an estimate of possible suppression 
 effects by 
 cosmological or $R^2$-terms; (2) a computation of the 
 total ADM energy of the zero modes; (3) a clarification (in 
 the Lorentzian case) of the influence of matter fields on 
 the fluctuations, with possible anomalous coupling. 
  
 Here, after a few general remarks about vacuum 
 fluctuations and ``spacetime foam" in quantum gravity, we 
 shall set out the general features of the dipolar fluctuations 
 (Section 1) and give some explicit order of magnitude 
 calculations (Section 2). Then we shall show that a 
 $\Lambda$-term cuts, to some extent, the dipolar 
 fluctuations (Section 3); this can lead in certain cases to an 
 anomalous coupling to matter (Section 4). In conclusion, 
 we shall discuss a number of new topics not addressed in 
 Ref.s \cite{npb,prd}. 
  
 \section{General features of the dipolar fluctuations} 
  
 The functional integral of pure Einstein quantum gravity 
 can be written as $z = \int d [g_{\mu \nu}] 
 \exp(i S/ \hbar)$, with 
 $S = \int d^4x \sqrt{g(x)} R (x)$. 
  
 The ``spacetime foam" \cite{foam} consists of fluctuations 
 whose action does not exceed a quantity of order $\hbar$. 
 This implies, for curvature fluctuations on a scale $d$, that   
 $| R | < G/d^4$ 
 (according to naive power counting) or 
 $| R | < 1/(L_{P}d)$  
 (according to numerical lattice estimates \cite{hw}). 
 Therefore, large fluctuations are expected to take place 
 only at very small distances. 
  
 Since, however, the Einstein action is not positive
 definite, one can also expect some fluctuations due to 
 peculiar cancellations of distinct contributions to the 
 action, which are by themselves larger than $\hbar$. 
  
 In order to work this out explicitely, let us consider the 
 Einstein equations with an 
 {\it auxiliary} source $T_{\mu \nu}$: 
  \begin{equation} 
 R_{\mu \nu}(x) - \frac{1}{2} 
 g_ {\mu \nu}(x) R(x) = 
 -8 \pi G T_{\mu \nu}(x) 
 \label{uno} 
 \end{equation} 
  and their trace 
  \begin{equation} 
 R(x) = 8 \pi G g^{\mu \nu}(x) 
 T_{\mu \nu}(x) \equiv 8 \pi G 
 {\rm Tr} T (x) 
 \label{due} 
 \end{equation} 
  
 Then consider a solution $g_{\mu \nu}(x)$ of (\ref{uno}) 
 with any source satisfying the condition 
  \begin{equation} 
 \int d^4x \sqrt{g(x)} {\rm Tr} 
 T(x) = 0 
 \label{croce} 
 \end{equation} 
  In view of (\ref{due}), this metric has zero action. We 
 have constructed in this way a zero mode of the pure 
 Einstein action. The source is unphysical, but it is 
 ``forgotten" after obtaining the metric. Condition 
 (\ref{croce}) means in fact that it is a ``dipolar" source, 
 with a compensation between regions having positive and 
 negative mass-energy density. Since this auxiliary source 
 is used to construct a virtual field configuration, we shall 
 sometimes call it a ``virtual source". 
  
 \section{Explicit computation to order $G^2$} 
  
 We have then found the following ``recipe" for 
 constructing dipolar zero-modes of the pure Einstein 
 action: given any source with zero integral (condition 
 (\ref{croce})), one solves the Einstein equations and finds 
 the corresponding metric. 
  
 Note, however, that the condition on the source already 
 contains $g_{\mu \nu}(x)$; furthermore, exact solutions 
 are in general not known, and the approximation error 
 could be such that the corresponding error on $S$ is larger 
 than $\hbar$. In this case we could still imagine that the 
 zero-mode can be computed in principle, but we would 
 not have in practice any definite idea of its properties. 
 An explicit evaluation is therefore needed. To this end we 
 consider static sources, with some free parameters 
 (typically $m_+$, $m_-$ and their sizes), in the weak field 
 approximation. We have in this static case 
  \begin{equation} 
 S_{zero-mode} = - \frac{1}{2} 
 \int d^4x \sqrt{g(x)} g^{\mu \nu}(x) 
 T_{00}(x) 
 \end{equation} 
  
 Using the Feynman propagator one finds to first order in 
 $G$ (compare \cite{npb}) 
  \begin{equation} 
 h_{\mu \nu} ({\bf x}) = 2G 
 (2\eta_{\mu 0} \eta_{\nu 0} - 
 \eta_{\mu \nu} \eta_{00}) 
 \int d^3y \frac{T^{00}({\bf y})}{| 
 {\bf x} - {\bf y}|} \label{campo} 
 \end{equation} 
  It is straightforward to check from this expression that 
  $\sqrt{g(x)} g^{00}(x) = 1 + o(G^2)$. 
 This means that the action of the metric generated by a 
 static source is simply 
  \begin{equation} 
 S = - \frac{1}{2} \int d^4x 
 T_{00}({\bf x}) + o(G^2) 
 \end{equation} 
  and provided the integral of the mass-energy density 
 vanishes, the field action is of order $G^2$, i.e.\ 
 practically negligible, as shown in the following numerical 
 example. 
  
 Consider a static dipolar source which is adiabatically 
 switched on/off with a lifetime $\tau$ of the order of 1 
 $s$. Suppose that the spatial size $r$ of the source is of 
 the order of 1 $cm$ and the two masses $m_\pm$ are of 
 the order of $10^k$ $g$, i.e.\ $10^{37+k}$ $cm^{-1}$ in 
 natural units. Note that with this mass we have, for the 
 ratio between the Schwarzschild radius and $r$, 
 $r_{Schw.}/r \sim 10^{-29+k}$. This implies that we can 
 compute $h_{\mu \nu}(x)$ in the weak field 
 approximation, with negligible error. More precisely, the 
 residual of second order in $G$ is found to be 
 \begin{equation} 
 S_{zero-mode}^{order \ G^2} \sim 
 \tau \frac{G^2 m^2_\pm}{r^3} 
 \sim 10^{-20+3k} 
 \end{equation} 
  
 Therefore the field of a static virtual source of this 
 magnitude order, satisfying the condition $\int d^3x 
 T_{00}({\bf x}) = 0$, has negligible action even if $k=6$, 
 corresponding to an apparent matter fluctuation with 
 density $10^6$ $g/cm^3$. This should be compared to the 
 action of the corresponding ``monopolar fluctuation", namely 
 $S_{monopolar} = (1/2) \tau m 
 + o(G^2) \sim 10^{47+k}$.

 \section{A $\Lambda$-term in the action cuts-off the 
 fluctuations} 
  
 The most recent estimates of the Hubble constant support 
 a non-zero value of the cosmological constant of the order 
 of $\Lambda \sim 10^{-50}$ $cm^{-2}$. 
 The cosmological term in the gravitational action, to be 
 added to the pure Einstein term, is 
  \begin{equation} 
 S_\Lambda = \frac{\Lambda}{8 \pi G} 
 \int d^4x \sqrt{g(x)} 
 \end{equation} 
  It is possible to evaluate the contribution of the dipolar 
 fluctuations to this term. This is easier for fluctuations 
 with spherical symmetry, like those generated by virtual 
 sources having the shape of ``+/- shells" (compare 
 \cite{npb}). In this case one can use the exact 
 Schwarzschild metric outside the source and the 
 spherically symmetric Newtonian field inside it. To 
 leading order one finds that 
  \begin{displaymath} 
 \Delta S_\Lambda = \frac{\Lambda \tau}{8 \pi G} 
\frac{1}{2} \int_{source} d^3x 
 {\rm Tr} h({\bf x}) = 
 \end{displaymath} 
 \begin{equation} 
  =\frac{\Lambda \tau}{4 \pi G} \int_{source} 
 d^3x V_{Newt.}({\bf x}) = 
 \Lambda \tau m r^2 Q 
 \end{equation} 
  where $Q$ is an adimensional factor which can be 
 negative or positive, depending on the distribution of the 
 positive and negative mass inside the virtual source. 
  
 Inserting for $\tau$, $m$ and $r$ the same values as 
 before, we find $\Delta S_\Lambda \sim 10^{-3+k}$. 
 Remembering that the lower bound for pure gravity was 
 $k \sim 6$, we see that the cosmological term cuts, to 
 some extent, the dipolar fluctuations. This works even 
 better for larger values of $r$.

\section{Matter coupling vs.\ local changes in $\Lambda$}

Let us consider now a scalar field $\phi$ coupled to gravity
	\begin{equation}
	L = \frac{1}{2} \left( \partial_\alpha \phi
	\partial^\alpha \phi - m^2 \phi^2 \right) 
\end{equation}
	\begin{equation}
	T_{\mu \nu} = \Pi_\mu \phi
	\partial_\nu \phi - g_{\mu \nu} L =
	\partial_\mu \partial_\nu \phi - g_{\mu \nu} L
\end{equation}
	\begin{equation}
	S_{interaction} = \frac{1}{2} \int d^4x
	 \sqrt{g(x)} T^{\mu \nu}(x) h_{\mu \nu}(x)
\end{equation}

To lowest order in $h_{\mu \nu}$ the interaction action can be
rewritten as
	\begin{equation}
	S_{interaction} = \frac{1}{2} \int d^4x \left(
	h_{\mu \nu} \partial^\mu \phi \partial^\nu \phi
	- {\rm Tr}h  L \right)
\end{equation}

On the other hand, the cosmological action is, still
to lowest order in $h_{\mu \nu}$ and
expanding $\sqrt{g} = 1 + \frac{1}{2} {\rm Tr}h + ...$
	\begin{equation}
	S_\Lambda = \frac{\Lambda}{8\pi G} \int d^4x
	\left( 1 + \frac{1}{2} {\rm Tr}h \right)
\end{equation}

Therefore the sum of the two terms can be rewritten as
	\begin{equation}
	S_{interaction} + S_\Lambda = \frac{1}{2} \int d^4x
	h_{\mu \nu} \partial^\mu \phi \partial^\nu \phi
	+ \frac{1}{2} \int d^4x  {\rm Tr}h
	\left( \frac{\Lambda}{8\pi G} - L \right)
\end{equation}

We see that to leading order the coupling of gravity to
$\phi$ gives a typical source term $(h_{\mu \nu} 
\partial^\mu \phi \partial^\nu \phi)$ and subtracts from
$\Lambda$ the local density $8\pi GL(x)$.
This separation is arbitrary, but useful and reasonable if the
lagrangian density is such to affect locally the
``natural" cosmological term and change the spectrum of 
gravitational vacuum
fluctuations corresponding to virtual mass densities
{\it much larger than the real density of} $\phi$.

Let us give an example. Suppose that $\phi$
represents a coherent fluid with the density of
ordinary matter ($\sim 1 \ g/cm^3$). At
the scale of 1 $cm$, with the observed value of 
$\Lambda$, the lower bound on virtual source density is
$\sim 10^3 \ g/cm^3$, which is much larger than the
real density. If $L$ is comparable to
$\Lambda/8\pi G$ in some region, an inhomogeneity in the 
cut-off mechanism of the dipolar fluctuations will follow, 
and this effect could dominate the effects of the
coupling $(h_{\mu \nu} \partial_\mu \phi
\partial_\nu \phi)$ to real matter.

 In our opinion, this dynamical mechanism could be the 
 basis for an explanation of the weak gravitational 
 modification by superconducting spinning disks which E.\ 
 Podkletnov claimed to have observed under very special 
 conditions \cite{ep,bos,umm}. A NASA/Argonne experimental 
 team is at work to replicate the original results, which at 
 this time are neither confirmed nor confuted. 
  
 From the very beginning of our theoretical analysis 
 \cite{shi} we maintained that the disks described in Ref.\ 
 \cite{ep} cannot be a source of gravitational field or 
 perturbate it {\it in a classical sense}. This is because the 
 gravitational coupling to matter is far too small, and 
 neither the presence of Cooper pairs inside the disk nor its 
 fast rotation help much under this respect. We have then 
 been looking for an interaction process not constrained by 
 the coupling, and a candidate for this are large quantum 
 fluctuations (compare our phenomenological model in 
 \cite{nat}).

More recently, anomalous gravity changes have been
observed during a solar eclipse \cite{cinesi}. In this case the
coherent matter which couples to the gravitational
fluctuations could be the native iron which is abundant
on the Moon.
  
 \section{Some remarks and new topics} 
  
 \medskip \noindent
 {\it (1) The fluctuations maintain $\langle 
 T_{\mu \nu}(x) \rangle = 0$.} 
  
 Even in the presence of strong quantum fluctuations, we 
 expect that the vacuum average of $T_{\mu \nu}(x)$, 
 computed through the functional integral, is zero. This is 
 defined as the average of $\frac{1}{8 \pi G} (R_{\mu \nu} 
 - \frac{1}{2} g_{\mu \nu} R)$. 
 More physically, there must be no apparent mass-energy-
 momentum density, on the average, in the vacuum. 
  
 An analogous property holds in QED. Even in the 
 presence of vacuum fluctuations, the vacuum average 
 value of the four-current, defined through Maxwell's 
 equations as $j_\mu = \partial^\nu F_{\mu \nu}$, 
 vanishes. 
  
 In fact, some general properties of the functional integral 
 ensure that the condition above on $T_{\mu \nu}$ is 
 respected, without any need of restricting the integration 
 space \cite{col}. In other words, the fluctuations 
 always average out in such a way to give a zero total 
 virtual mass density at any point. 
  
 \medskip \noindent
 {\it (2) The dipolar zero-modes have $S=0$ but are not 
 extrema of $S$.} 
  
 The dipolar zero-modes are not a minimum of the pure 
 Einstein action, because this would be equivalent of being 
 a solution of the vacuum Einstein equations; but the 
 dipolar zero-modes do not satisfy the Einstein equations in 
 vacuum, even through they have the same action as the 
 vacuum solutions ($S=0$). 
  
 We can understand better the situation at hand through a 
 bidimensional analogue (while, of course, the space of the 
 possible metric configurations is infinite-dimensional). Let 
 us consider the function $f(x, y) = x^2 y^2$ and draw its 
 cartesian plot. We find that it resembles a paraboloid, 
 except for the fact that the axes $x=0$ and $y=0$ are a 
 sort of ``cuts"; along these axes the function takes the value 
 $f=0$, which is also the value at the origin. However, the 
 origin is a minimum, while the points on the axes distinct 
 from the origin are not minima. 
  
 \medskip \noindent
 {\it (3) How much is the phase space volume of dipolar 
 fluctuations?} 
  
 The example above suggests another possible property of 
 the dipolar zero-modes: in the same way as the ``zero 
 lines" of the function above have null measure with 
 respect to its full bidimensional domain, one may think 
 that the condition for the dipolar zero-modes defines a 
 subspace of all field configurations having lower 
 dimension and thus null measure. In this case the dipolar 
 fluctuations would be suppressed because they have zero 
 volume in phase space. 
  
 This would be true if the dipole condition $m_+ = m_-$ 
 had to be satisfied exactly. In fact, however, we 
 considered weaker conditions. We saw, for instance, that 
 up to values of the virtual mass of the order of $10^6$ $g$ 
 the difference between $m_+$ and $m_-$, i.e., the width 
 of the ``line in phase space", is of order $\tau^{-1}$, where 
 $\tau$ is the duration of the fluctuation. 
 
 The appearance of 
 $\tau$ in this relation could also provide an upper bound 
 on the duration of the fluctuations, making $\tau$ closer to 
 the minimum duration allowed by the Heisenberg 
 principle, like it happens for electromagnetic or scalar 
 fields (compare \cite{npb,prd}). 
 In other words, a long lifetime $\tau$ requires a very 
 precise compensation between the positive and negative 
 masses of the virtual dipole, thus implying a very small 
 phase space volume for that configuration. 
  
 \medskip \noindent
 {\it (4) Strong dipolar fluctuations are not allowed in 
 QED.} 
  
 This is an immediate consequence of the quadratic form of 
 the electromagnetic lagrangian density: $L \propto ({\bf 
 E}^2 - {\bf B}^2)$. The contribution to the action of the 
 +/- charges in a virtual dipole is the same, therefore there 
 are no cancellations. The dipolar fluctuations are not 
 favoured with respect to the monopolar fluctuations. 
 
 The 
 reasoning above also leads to the following statement. 
  
 \medskip \noindent
 {\it (5) The null-action property of the dipolar fields is a 
 non-perturbative feature.} 
  
 In fact, suppose we limit ourselves to consider, as usual in 
 perturbation theory, the part of the action quadratic in $h$ 
 in an expansion around the minimum. We then would be 
 in the same situation as for the electromagnetic field. It is 
 easy to see that the quadratic part of the gravitational 
 action is not positive definite and has positive and 
 negative eigenvalues in the Euclidean formulation, unlike 
 the electromagnetic action \cite{prd59}. Nevertheless, any 
 single term would be the same for positive or negative 
 virtual sources (see eq.\ (\ref{campo})), without any 
 cancellation. 
  
 In other words, the null action property of the dipolar zero 
 modes cannot be obtained just considering the expansion 
 in powers of $h$. This not because we are in strong field 
 conditions, but because the zero-modes are not local 
 minima of the action (compare Point 2). 
  
 \medskip \noindent
 {\it (6) The $T_{\mu \nu}$ of any real pure 
 electromagnetic field generates gravitational zero-modes,
 up to terms of order $G^2$.} 
  
 Since the trace of the energy-momentum tensor of any 
 electromagnetic field is zero, it is possible to use, instead 
 of virtual unphysical dipolar mass sources, a {\it real} 
 electromagnetic field as source of gravitational zero-
 modes. It needs to be a pure electromagnetic field in 
 vacuum, thus any plane waves or wave packets are 
 admissible. We stress again, however, that the real or 
 virtual character of the source is irrelevant in order to 
 obtain vacuum fluctuations. 
  
 \bigskip \noindent {\bf Acknowledgments} - This work was
 supported in part by the California Institute for Physics
 and Astrophysics via grant CIPA-MG7099. The author is
 grateful to C.\ Van Den Broeck, M.\ Gross and J.\
 Brandenburg for useful discussions and remarks.

 \end{document}